\documentclass[twocolumn,prl,showpacs]{revtex4}
\usepackage{graphicx}
\usepackage{amsmath}
\usepackage{amssymb}
\usepackage[ansinew]{inputenc}

\begin{document}
\title{The Luttinger model following a sudden interaction switch-on}
\author{M.~A. Cazalilla}
\affiliation{Donostia International Physics Center (DIPC), 
Manuel de Lardizabal 4, 20018-Donostia, Spain}
\begin{abstract} 
The evolution of correlations in the \emph{exactly} solvable Luttinger model (a model of interacting  fermions in one dimension) after a sudden interaction switch-on  is \emph{analytically} studied.  When the model is defined on a finite-size ring, zero-temperature correlations are periodic in time. However, 
in the thermodynamic limit, the system relaxes algebraically towards a stationary state which is well described, at least for some simple correlation functions, by the generalized Gibbs ensemble recently introduced by Rigol \emph{et al.} [cond-mat/0604476].  The critical exponent that characterizes the decay of the one-particle correlation function is different from the known equilibrium exponents.  Experiments for which these results can be relevant are also discussed. 
\end{abstract}

\maketitle

 Experiments with cold atomic gases are motivating research into problems
 that, previously,  would have looked  highly academic. One such problem concerns the evolution of a quantum many-body system where  interactions (or other parameters of the system) are time-dependent. An example is an interaction \emph{quench}: an experiment where  the strength  of interactions is suddenly changed. This type of experiment is  nowadays feasible thanks to the phenomenon known as Feshbach resonance~\cite{RCBGCW98,LRTBJ02}, which allows to tune the strength and sign of interactions in a cold atomic gas by means of a magnetic field. If the applied magnetic field is time dependent, the interactions become time-dependent.  Alternatively, in optical lattices~\cite{GMEHB02}, it is possible to change the lattice parameters  in a time-dependent fashion, which effectively amounts to varying the ratio of the interaction to the kinetic energy in time. On the theory side, the recent development of extensions of the density-matrix renormalization group (DMRG) algorithm~\cite{CM02,V04,DKSV04,WF04,VGC04}  has spurred  the interest in understanding the properties of quantum many-body systems out of equilibrium and, in particular, in the dynamics following a quench.

  Because of these new possibilities, the evolution of observables  and correlations following a sudden change of the system parameters is attracting 
 much theoretical  interest~\cite{AA02,BL05,YAKE05,KSDZ05,AV05,RCM06,YD06,CC06,RDYO06,RMRNM06}.  One interesting question that has been raised  by a recent 
 experiment in an array of 1D cold atomic gases~\cite{KWW06}
 is whether after a quench a system possessing an infinite number of integrals of
 motion can exhibit relaxation towards a steady state or not.
 This question has been analyzed by the authors of Ref.~\cite{RDYO06},
 who have \emph{numerically} shown that the steady state 
 of an integrable gas of hard-core
 bosons   is described by a generalized Gibbs distribution that maximizes
 the entropy  with all possible constraints imposed by the existence
 of the (infinite number of) integrals of motion. 
 Here the effect of suddenly turning on the interactions in the 
 Luttinger model is \emph{analytically} studied. It is  shown that, when the model is defined on a finite-size ring, the asymptotic form of the two-point one-body and density correlations   at zero temperature  is periodic in time, and therefore the system exhibits no relaxation to a steady state with time-independent properties. In the termodynamic limit, however, the  same  correlation functions relax to a steady state, whose properties are different from  those of the ground state. Indeed, the decay of the   one-particle correlations  with distance is governed by a critical 
exponent which is different from the known equilibrium exponents. 
Interestingly,  one-particle and density correlations in the steady state 
 can be obtained  using the generalized Gibbs ensemble introduced 
 by Rigol \emph{et al.} in Ref.~\cite{RDYO06}.
  
 The Luttinger model  (LM) describes a system  of interacting Fermions in one
 dimension (1D). It was introduced by Luttinger~\cite{L63} in 1963, but the correct   exact  solution was  found in 1965  by Mattis and Lieb~\cite{LM65}.  Asymptotic forms of one and two-particle correlations in equilibrium were  obtained by Luther and Peschel~\cite{LP74}. Later, 
 Haldane~\cite{H80,H81a,H81b} proposed that this model  
 describes the low-energy properties 
 of a fairly broad class of  systems in 1D known as 
 Tomonaga-Luttinger  liquids~\cite{H80,GNT98,G04}.

 The Hamiltonian of the LM,  $H_{LM} = H_0 + H_2 + H_4$, where
$H_0 =  \sum_{p,\alpha}  \hbar v_F p \, : \psi^{\dag}_{\alpha}(p) \psi_{\alpha}(p):$ is the free-fermion Hamiltonian, and the interactions are described by
\begin{eqnarray}
H_{2} &=& \frac{2\hbar\pi}{L} \sum_{q} g_2(q) \, J_R(q) J_L(q), \\
 H_{4} &=& \frac{\hbar \pi}{L} \sum_{q,\alpha} g_4(q) \: :J_{\alpha}(q) J_{\alpha}(-q): \,\, ,
\end{eqnarray}
where the Fermi operators $\{\psi_{\alpha}(p), \psi^{\dag}_{\beta}(p')\} =  \delta_{p,p'} \delta_{\alpha,\beta}$ ($\alpha, \beta =   L,R$) and anti-commute otherwise. To avoid a degenerate ground state, anti-periodic
boundary conditions are chosen: $\psi_{\alpha}(x + L) = - \psi_{\alpha}(x)$,
($\psi_{\alpha}(x) =\sum_{p} e^{i s_{\alpha} p x} \: \psi_{\alpha}(p)/\sqrt{L}$
is the Fermi field operator  and $s_{R} = -s_{L} = +1$ and $L$ the length
of the system) so that $p = 2\pi (n - \frac{1}{2})/L$, and $n$ is an
integer. The ``current'' operators $J_{\alpha}(q) = \sum_{q} \, 
:\psi^{\dag}_{\alpha}(p+q) \psi_{\alpha}(p):$, where $q = 2\pi m/L$, 
$m$ being an integer; $:\ldots:$ stands for the normal order prescription
according to which all creation operators are to be found to the left of the annihilation operators and expectation values over the ground state of $H_0$
are subtracted.  Thus, the above model describes a system of fermions 
interacting via the four Fermion terms $H_2$ and $H_4$.
Fermions come in two chiralities, $R$ standing for right moving and $L$ for
left moving particles, respectively.  The dispersion is linear  and therefore it is not bounded from below. To define a stable ground state for $H_0$ 
all single-particle  levels with $p < 0$ are filled up for both chiralities, which yields a Dirac sea  (\emph{i.e.} an ``infinite story'' hotel) which will be
denoted as $|0\rangle$.  The coupling functions $g_2(q)$ and $g_4(q)$
are assumed to be finite for $q = 0$. Moreover, 
to ensure that the Hilbert space of $H_{LM}$ and $H_0$ remain the same
and, in particular, that their ground states have a finite overlap at 
finite $L$, $g_2(q)/(v_F + g_4(q)) \to 0$ faster than $|q|^{-1/2}$
as $|q| \to \infty$ and $|g_2(q)| <  v_F + g_4(q)$ for all $q$~\cite{H81a}.

  The currents obey a Kac-Moody algebra~\cite{H81a,GNT98,G04}:
$[J_{\alpha}(q), J_{\beta}(q')] = \frac{qL}{2\pi}\delta_{q+q',0} \:
\delta_{\alpha,\beta}$. This fact allows to introduce, for $q \neq 0$, the following
operators  $b_0(q) = -i \left( 2\pi/|q|L \right)^{1/2} \left[   \theta(q) 
J_{R}(-q) - \theta(-q) J_{L}(q) \right]$ and 
$b^{\dag}_0(q) = i \left( 2\pi/|q|L \right)^{1/2} \left[   \theta(q) 
J_{R}(q) - \theta(-q) J_{L}(-q) \right]$, which obey the standard algebra of 
boson (``phonon'') operators. Moreover, there are two conserved
operators $\delta N = N_{R} + N_{L}$, {\emph{i.e.}} the number operator
referred to the ground state $|0\rangle$, and the total current 
 $J = N_{R} - N_{L}$, where $N_{\alpha} = J_{\alpha}(0)$. For fermions,
 the physical states obey the selection rule $(-1)^{\delta N}
 = (-1)^J$. In terms of the boson operators $b^{\dag}_0(q), b_0(q)$
 the Hamiltonian $H_{LM}$ is quadratic but not diagonal. It can
 be diagonalized by means of a Bogoliubov  (`squeezing') 
 transformation~\cite{LM65}: 
\begin{eqnarray}
b(q) &=& \cosh \varphi(q) \: b_0(q) + \sinh \varphi(q) \: b^{\dag}_{0}(-q), 
\label{eq:bogol1}\\
b^{\dag}(q) &=&  \sinh \varphi(q)\:  b_0(-q) + \cosh \varphi(q)\:  b^{\dag}_0(q).\label{eq:bogol2}
\end{eqnarray}
To render $H_{LM}$ diagonal, we must choose $\tanh 2\varphi(q) = g_2(q)/[v_F + g_4(q)]$. Thus the Hamiltonian becomes~\cite{H81a} 
$H_{LM} = \sum_{q\neq 0} \hbar v(q) |q| b^{\dag}(q) b(q) + \hbar\pi v_N\: \delta N^2/L  + \hbar \pi v_J \: J^2/L$, where $v(q) = [(v_F + g_4(q))^{2} - g^2_2(q)]^{1/2}$,  $v_N = v e^{2\varphi}$, and~\cite{H81a} 
$v_J = v e^{-2\varphi}$, being $v = v(0)$ and
$\varphi = \varphi(0)$.

  Let us now consider an interaction quench in the LM. Here I consider only
the case where the coupling functions $g_2(q)$ and $g_4(q)$ are suddenly
switched on at $t = 0$. Thus, the initial state of the system will be described by
a thermal distribution determined by  the non-interacting Hamiltonian $H_0$,
$\rho(t =0) = \rho_{0} =  e^{-H_0/T}/Z_0$, 
where $Z_0 = {\rm Tr} \: e^{-H_0/T}$. However, for $t > 0$, the evolution is
dictated by the full Hamiltonian $H_{LM}$. A more general type of quench
corresponds to a sudden switch between two different forms of $g_2(q)$ and
$g_4(q)$. Whereas the results described below can be generalized to such a
case, I believe a quench from the non-interacting limit is most interesting because the spectrum
of $H_0$ contains free fermions whereas the spectrum of $H_{LM}$ 
does not~\cite{LP74,H81a,GNT98,G04}. Thus, a sudden switch-on of the interactions describes a time-dependent destruction of the characteristic discontinuity of the momentum distribution at the Fermi points $p = 0$.  

 Equal time correlations of a given operator $O(x)$,
\begin{eqnarray}
C_{O}(x,t) &=& \langle e^{i H_{LM} t/\hbar} O^{\dag}(x) O(0) e^{-iH_{LM} t/\hbar} \rangle_0 \\
 &=& {\rm Tr} \: \rho_0\,  e^{i H_{LM} t/\hbar} O^{\dag}(x) O(0) e^{-i H_{LM} t/\hbar}.   \label{eq:correlation}
\end{eqnarray}
Note that since $[H_0, H_{LM}] \neq 0$, $C_{O}(x,t)$ is explicitly time-dependent. Indeed, in the LM model time dependence stems from $H_2$, since $[H_0,H_4] = 0$. $H_2$ describes scattering between fermions moving in opposite directions,  and, as shown below, it produces entanglement between the excitation modes with $q >0$ and $q < 0$. 

 The exact evolution of  $b_0(q)$  has a fairly simple form:
 \begin{equation}
 b_0(q,t) =  f(q,t) b_0(q) + g^{*}(q,t) b^{\dag}_0(-q), \label{eq:solution}
 \end{equation}
 where $b_0(q,t) = e^{iH_{LM}t/\hbar} b_0(q) e^{-i H_{LM}t/\hbar} $, 
 $f(q,t) = \cos v(q) |q| t   - i \sin v(q) |q| t \cosh 2\varphi(q)$, and
 $g(q,t) = i \sin v(q) |q| t \sinh 2 \varphi(q)$. Note that this form obeys the
 correct boundary condition $b_0(q,0) = b_0(q)$. Entanglement between
modes of opposite $q$ vanishes for $\varphi(q) = 0$ (\emph{i.e.}
$g_2(q) = 0$)  in agreement with the above discussion. 

  The evolution of  one-particle correlations (\emph{i.e.} $O(x) = \psi_{\alpha}(x)$) can be obtained from Eq.~(\ref{eq:solution}) and the bosonization formula~\cite{LP74,H81a,GNT98,G04}:
\begin{equation}
\psi_{\alpha}(x) = \frac{\eta_{\alpha}}{(2\pi a)^{1/2}} \,  e^{i s_{\alpha} \phi_{\alpha}(x)},\label{eq:bosonization}
\end{equation}
being $\eta_{R} \neq \eta_{L}$ two different Pauli matrices that ensure the anti-commutation of the left and right-moving Fermi fields;  $\phi_{\alpha}(x) = s_{\alpha} \varphi_{0\alpha} + 2\pi x N_{\alpha}/L + \Phi^{\dag}_{\alpha}(x) + \Phi_{\alpha}(x)$, where $[N_{\alpha},\varphi_{0\beta}] = i \delta_{\alpha,\beta}$,
$\Phi_{\alpha}(x) = \sum_{q > 0} \left(2\pi/qL \right)^{1/2} e^{-qa/2} e^{iqx} \:
b^{\dag}_0(q)$, and $a \to 0^{+}$. Setting 
$\sinh 2\varphi(q) = e^{-|q R_0|/2} \sinh \varphi$, where $R_0 \ll L$ is of the order  of the range of the interactions,  and replacing $v(q)$ by 
its $q = 0$ value~\cite{LP74}, simplifies the calculations without 
altering the asymptotic form of the correlations. At $T = 0$, for a system 
of size $L$,  one-body correlations are given by the following
expression~\footnote{The correlation function
for $\psi_L(x)$  can be obtained by  replacing $x$ by $-x$.}:
\begin{eqnarray}
C_{\psi_R}(x,t > 0|L)  = G^{(0)}_R(x|L) \left[\frac{ R_0}{d(x|L)} \right]^{\gamma^2} \nonumber \\
\times
 \left[ \frac{d(x - 2v t|L) d(x + 2vt |L)} {[d(2 vt|L)]^2} \right]^{\gamma^2/2}, 
\label{eq:oneparticle}
\end{eqnarray}
where $d(x|L) = L |\sin(\pi x/L)|/\pi$ is the \emph{cord} function,  
$G^{(0)}_{R}(x|L) = i/[2 L \sin\pi(x + i a)/L]$ the 
non-interacing correlation function,
and $\gamma = \sinh 2 \varphi$. The above  expression is accurate asymptotically, \emph{i.e.} for $d(x\pm 2vt|L), d(x|L), d(2 vt|L) \gg R_0$. 
It can be seen that the one-particle correlations are periodic in time:
$C_{\psi_{R}}(x, t+ T_0|L) = C_{\psi_{R}}(x,t|L)$ with
$T_0 = L/2v$. This implies that the finite-size LM does not relax, which is 
a consequence of the (approximately) linear dispersion of the eigenmodes
near $q = 0$ along with the absence of any damping mechanisms in the LM
(see discussion at the end).  However, in the thermodynamic limit, $L \to \infty$, and $d(x|L) \to |x|$.  Therefore, Eq.~(\ref{eq:oneparticle}) becomes:
\begin{equation}
C_{\psi_R}(x,t > 0) = \frac{i}{2\pi(x + i a)} \left| \frac{R_0}{x}\right|^{\gamma^2} 
\left| \frac{x^2 - (2v t)^2}{(2vt)^2} \right|^{\gamma^2/2}. \label{eq:onebodyth}
\end{equation}
It is interesting to analyze the above expression in the limit where $2 vt \ll |x|$,
where  it becomes
\begin{equation}
C_{\psi_R}(R_0 \ll 2v t \ll |x| ) \approx \frac{iZ(t)}{2\pi (x + i a)},
\end{equation}
being $Z(t) = (R_0/2vt)^{\gamma^2}$ a time-dependent renormalization
constant of the Fermi quasi-particles. Thus for short-times the system behaves
as a Femi liquid, with a singularity at  the Fermi points given by $Z(t)$, which 
decreases with time. On the other hand, for $2 vt \gg |x|$ 
the correlation takes a non-Fermi liquid form:
\begin{equation}
C_{\psi_R}(R_0 \ll  |x| \ll 2 vt ) \approx \frac{i}{2\pi(x + i a)} \left| \frac{R_0}{x}\right|^{\gamma^2} \label{eq:relaxed1p}
\end{equation}
In particular, in the limit $t \to +\infty$  one-particle correlations relax to the 
power-law in the right-hand side of Eq.~(\ref{eq:relaxed1p}). Notice that, 
although $C_{\psi_R}(x,t \to \infty)$ exhibits a power-law behavior, the latter
is governed by an exponent that is different from the one that governs
asymptotic ground-state correlations~\cite{LP74,H81a},  $\gamma^2_0 = 2 \sinh^2 \varphi <  \gamma^2 = \sinh^2 2 \varphi$ for $\varphi \neq 0$. 
The origin of this new exponent will be discussed below.

The different behavior of $C_{\psi_R}(x,t)$ for short and long times  can be understood in terms of a `light-cone' effect~\cite{CC06}: The initial state $|0\rangle$ has higher energy than the ground 
state of $H_{LM}$ (see discussion further below). Therefore, it contains long
wave-length  phonons that propagate from time $= 0$ to time $= t$ along
light-cones where the role of speed of light is played by $v$. These  excitations determine which points retain the same type of correlations found in $|0\rangle$ and which  points acquire new correlations. 
The latter phenomenon and the overall structure of (\ref{eq:onebodyth}) bears some resemblance to results reported in Ref.~\cite{CC06}. Nevertheless, I have so far failed to extend the methods of~\cite{CC06} to the quench in the LM. 
There are two main  differences: the initial state in the present case is
non-critical, and therefore it does not have any characteristic (gap) energy scale  as the initial states considered in~\cite{CC06}. Secondly, and more  importantly, the critical exponent found above is different from
the bulk or boundary exponents of the field operator $\psi_R(x)$. Indeed, this
may be an indication that the quench in the LM belongs to a different
universality class.

  One may think that the relaxation behavior exhibited by $C_{\psi_R}(x,t)$
 in the thermodynamic limit is because the field operator, 
 $\psi_R(x)$, is a non-linear function of $b_0(q)$ and $b^{\dag}_0(q)$. 
 However, the (density) operator $J_{R}(x) = \partial_x \phi_R(x)/2\pi$
 also exhibits relaxation. Setting $O(x) = J_R(x)$ in (\ref{eq:correlation}), the following  is obtained using Eq.~(\ref{eq:solution}),
 \begin{eqnarray}
 C_{J_R}(x,t|L) &=& -\frac{1}{4\pi^2}\Big\{ \frac{1 + \gamma^2}{[d(x|L)]^2} \nonumber \\   -&&  \frac{\gamma^2}{2[d(x - 2vt)]^2} - 
 \frac{\gamma^2}{2[d(x + 2vt)]^2} \Big\}.
 \end{eqnarray}
For finite $L$ the density correlation function is again periodic in time. However,
for $L \to \infty$, it shows relaxation:
$C_{J_R}(x, t \to \infty| L)  \to -  (1+\gamma^2)/(4 \pi^2 x^2)$. This form
again deviates from the ground state behavior, where the prefactor of $-1/(4 \pi^2 x^2)$ is $\cosh 2\varphi -\sinh 2\varphi = e^{-2\varphi}$~\cite{H81a,GNT98,G04}. 

 It is interesting to find that the above results in the $t \to \infty$ limit can be analytically obtained from the generalized Gibbs distribution introduced in
 Ref.~\cite{RDYO06}, which is described by the following density  matrix:
 \begin{equation}
 \rho_{gG} = \frac{1}{Z_{gG}} e^{\sum \lambda(q) I(q)},
 \end{equation}
where $Z_{gG} = {\rm Tr}\:  e^{\sum \lambda(q) I(q)}$ and $[H, I(q)]  = [I(q), I(q')] = 0$, that is, a set of independent integrals of motion. Since $[H_{LM}, n(q)] =0$, where $n(q) = b^{\dag}(q) b(q)$, the  phonon occupancy operators seem as the most natural choice for $I(q)$. The Lagrange multipliers $\lambda(q)$ are obtained from the  condition~\cite{RDYO06}:
\begin{equation}
\langle n(q) \rangle_{t = 0} = \langle 0| n(q) | 0\rangle = \langle n(q) \rangle _{gG}
= {\rm Tr} \: \left[ \rho_{gG} n(q) \right],
\end{equation} 
where $T = 0$ was assumed. Using (\ref{eq:bogol1},\ref{eq:bogol2}), 
$\langle 0 | n(q) | 0 \rangle = \sinh^2 \varphi(q)$, which is a non-thermal
distribution. However,  $\lambda(q)$ do not need to obtained explicitly, as 
it suffices to realize that $\rho_{gG}$ has the same form as the
distribution in the canonical ensemble with $H/T = -\sum_{q} \lambda(q) n(q)$. 
One can also regard $\rho_{gG}$ as  a canonical
distribution with a $q$-dependent
temperature, $T(q) = - \hbar v(q) |q| /\lambda(q)$.  Using this fact along with
equations~(\ref{eq:bogol1},\ref{eq:bogol2}) and (\ref{eq:bosonization}),
I find that
\begin{eqnarray}
C^{gG}_{\psi_R}(x) = {\rm Tr} \: \rho_{gG} \: \psi^{\dag}_{R}(x) \psi_{R}(0) = \lim_{t \to +\infty}
C_{\psi_R}(x,t), \label{eq:res1}\\
C^{gG}_{J_R}(x) = {\rm Tr}\: \rho_{gG}\: J_R(x) J_R(0)  = \lim_{t \to +\infty}
C_{J_R}(x,t).\label{eq:res2}
\end{eqnarray}
Thus, at least for these simple correlation functions, it seems that the 
generalized Gibbs distribution describes the stationary state of the LM 
after an interaction quench. The reason why the critical
exponent $\gamma^2$ turns out to be different from the known equilibrium
exponents can  be thus explained in two different ways: mathematically,
it is seen that in order to obtain the evolution of the operator $b_0(q)$,
Eq~(\ref{eq:solution}), one has to do and undo the Bogoliubov 
transformation (\ref{eq:bogol1},\ref{eq:bogol2}). However, these
transformations do not cancel
each other exactly (except at $t = 0$) because of the  phase
factors $e^{\pm i v(q) |q| t}$ introduced by the time evolution operator. In contrast, in the equilibrium problem, 
since the expectation value is taken over the ground state of $H_{LM}$,
the Bogoliubov transformation is performed only once. Physically,
in view of the results~(\ref{eq:res1},\ref{eq:res2}), the difference in exponent 
can be regarded a consequence of  
the non-equilibrium distribution of phonons $\langle 0| n(q)|0 \rangle = 
\sinh^2 \varphi(q)$, which is a constant of motion. 
Note as well that  $\langle 0| n(q) n(q')   |0 \rangle  - \langle 0| n(q) | 0\rangle
\langle 0| n(q') |0 \rangle = \left[ \sinh^4 \varphi(q) + 2 \cosh^2 \varphi(q) 
\sinh^2 \varphi(q) \right] \delta_{q,q'}$ is non-zero for $q=q'$,
since $|0\rangle$ is not an eigenstate of $H_{LM}$ for $g_2(q) \neq 0$.

  Let us finally consider how the above predictions could be  experimentally 
observed. To date, there are no exact realizations of the LM in nature. However,
one can exploit the fact that the LM describes the low-energy properties
of  Tomonaga-Luttinger liquids~\cite{H81a,GNT98,G04}, of which
several physical realizations  in cold atomic 
gases are available~\cite{KWW06,SMSKE04,GSMKE05}. 
Let us therefore consider a single-species cold Fermi gas confined
to 1D in a strongly anisotropic trap~\cite{GSMKE05}.  In a single-species
cold Fermi gas, the p-wave interaction is naturally negligible. 
One possibility to realize a 
sudden change of the interaction is to use a
p-wave Feshbach resonance~\cite{GSMKE05}, which enhances the
strength of this interaction.
Alternatively,  one can use a 1D
dipolar Fermi gas, where interactions at long distances
are described by the potential:
\begin{equation}\label{eq:dipole}
V_{\rm dip}(x) = \frac{1}{4\pi \epsilon_0} \frac{D^2 (1 - 3 \cos \theta)}{[x^2 + R^2_0]^{3/2}},
\end{equation}
 $D$ being the dipolar momentum of the atoms and $\theta$ is the angle subtended  by the direction of the atomic motion and an electric  field (or magnetic, for magnetic dipoles) that polarizes the gas.  In the above
expression $R_0$ is of the order of the transverse size of the cloud. 
The Fourier transform of (\ref{eq:dipole}), $g_2(q) = g_4(q) \propto 
V_{\rm dip}(q) = \lambda(\theta) |q R_0| K_1( |q R_0|)$, where 
$\lambda(\theta) = D^2 (1 - 3 \cos \theta)/2\pi \epsilon_0 R^2_0$ and $K_1(x)$
is the first order modified Bessel function. 
An sudden switch-on of $V_{\rm dip}(q)$ 
can be realized by  deviating the electric 
field   that polarizes the gas  from the 
``magic'' angle $\theta_m = 
\cos^{-1}(\frac{1}{3})$, for which the interaction vanishes 
(\emph{i.e.} $\lambda(\theta_m) = 0$).  

  However, the full Hamiltonian for a TLL contains an infinite series of 
terms that spoil the  integrability of the LM~\cite{H80,H81a}. Roughly 
speaking, these stem from the non-linearity of the fermion 
dispersion~\footnote{It is possible to extend the above calculations
to Fermions in a harmonic trap, where the dispersion is linear~\cite{unpub}.}
and the fact that interactions  couple right and left-moving modes in a 
way that is highly non-linear in terms of the boson fields $\phi_{\alpha}(x)$
(\emph{umklapp} scattering)~\cite{H80}. In a TLL all these deviations 
are irrelevant in the renormalization-group
sense, which means that their effect on low-energy states is small. 
Nevertheless,  after a sudden change of the interaction in the systems
described above, high-energy excitations
will be created that are not described by the LM.
Exciting many fermions to levels very far from the Fermi level 
where the LM description is not accurate 
can be avoided  by turning on the interaction to a value much
smaller than the Fermi energy. On the other hand, low
energy excitations will survive for longer times and, since
they dominate the long-time dynamics, the behavior of the correlations
will be described by the above results. Thus,
if the quench was conducted at zero temperature, since the 
atomic systems are finite, an approximately periodic behavior 
of correlations can be expected.
However, Fermi gases are usually hard to cool down,
and a situation where temperature is larger than level spacing
(\emph{i.e.} $T \gg \pi/L$) is perfectly reallistic. In this situation,
one should consider correlations at finite $T$, neglecting finite size effects.
The latter can be obtained from Eq.~(\ref{eq:oneparticle}) upon replacing
$L \sin(\pi x/L)/\pi$ by $(\hbar  v_F/\pi T) \sinh(\pi T x/\hbar v_F)$, etc. Thus relaxation 
takes place because temperature induces a finite
correlation length in the initial state and therefore correlations  decay
exponentially. One-body correlations that can be accessed  through the momentum distribution, which can be measured in a time of light experiment.
Thus the steady state momentum distribution
following a sudden switch-on of interactions 
should differ from the equilibrium distribution at the same tempereature. 
A more detailed analysis will be given elsewhere~\cite{unpub}.

 This work was supported by \emph{Gipuzkoako Foru Aldundia} and
MEC (Spain) under grant  FIS2004-06490-C03-00.
 \bibliography{quenchletter}
\end{document}